\documentclass{mn2e}
\usepackage{epsfig}
\def\msun{{\rm M_{\odot}}}

\def\me{{\dot M_{\rm Edd}}}

\def\le{{L_{\rm Edd}}}

\title[AGN Black Holes: Underweight and
  Eddington] {AGN have Underweight Black Holes and
  Reach Eddington}

\author[A. R. King] {A. R. King$^{1}$ \\ $^1$Theoretical Astrophysics
  Group, University of Leicester, Leicester LE1 7RH}

\date{\today}

\volume{000}

\pagerange{\pageref{firstpage}--\pageref{lastpage}} \pubyear{2005}

\begin{document}

\label{firstpage}

\maketitle

\begin{abstract}

Eddington outflows probably regulate the growth of supermassive black
holes (SMBH) in AGN. I show that effect of the Rayleigh--Taylor
instability on these outflows means that SMBH masses are likely to be a
factor of a few below the $M - \sigma$ relation in AGN. This agrees
with the suggestion by Batcheldor (2010) that the $M - \sigma$
relation defines an upper limit to the black hole mass. I further
argue that observed AGN black holes must spend much of their lives
accreting at the Eddington rate. This is already suggested by the low
observed AGN fraction amongst all galaxies despite the need to grow to
the masses required by the Soltan relation, and is reinforced by the
suggested low SMBH masses. Most importantly, this is the simplest
explanation of the recent discovery by Tombesi et al (2010a, b) of the
widespread incidence of massive ultrafast X--ray outflows in a large
sample of AGN.

\end{abstract}

\begin{keywords}
  accretion: accretion discs -- galaxies: formation -- galaxies:
  active -- black hole physics
 
\end{keywords}

\section{Introduction}

It is now well established that the centre of almost every galaxy
contains a supermassive black hole (SMBH), and that in many cases the
mass of this hole is closely connected with properties of the host
galaxy bulge through the $M - \sigma$ and $M -M_{\rm bulge}$ relations
(Ferrarese \& Merritt, 2000; Gebhardt et al. 2000; H\"aring \& Rix
2004). The connection is physically reasonable, since the black hole
binding energy $\eta Mc^2$ considerably exceeds the bulge binding
energy $\sim M_{\rm bulge}\sigma^2$ (King, 2003) (here $\eta\sim 0.1$
is the accretion efficiency and $\sigma$ the velocity dispersion of
the bulge).  The black hole communicates its presence to the host by
driving powerful outflows when it is fed matter at super--Eddington
rates. If these outflows are momentum--driven, i.e. communicate only
their ram pressure to the surrounding interstellar gas, the $M -
\sigma$ relation emerges naturally as specifying the black hole mass
at which an Eddington outflow can drive a significant bubble into the
bulge gas (King, 2003; 2005). At smaller masses, the black hole can
only drive bubbles which recollapse, and evidently do not interrupt
the gas supply to the black hole which ultimately powers its growth.

This argument implicitly suggests that the $M - \sigma$ relation is an
upper limit to the black hole mass, rather than a tight relation.  It
is now clear (Batcheldor, 2010) that this is probably so, as
observational selection makes it difficult to measure black hole
masses below the relation (cf Section 3).

The obvious question then is how far below this limit the majority of
SMBH lie.  New insight into this question comes from recent X--ray
observations by Tombesi et al. (2010a, b) of fast ($v \sim 0.1c$)
outflows in a large fraction of local AGN.  I shall argue here that
this means that most local AGN contain black holes lying below the $M
- \sigma$ limit, and that most of these systems undergo
super--Eddington episodes which switch off only for relatively short
intervals.  The hole masses are probably not very far below the $M -
\sigma$ value. I will show that Eddington outflows at masses
significantly below this are Rayleigh--Taylor unstable and therefore
inefficient in suppressing accretion.

\section{Black Hole Growth}

The argument by Soltan (1982) relates the mass density of black holes
in the local universe to the total background radiation they produced
while growing. It suggests that the average medium--to--large galaxy
hosts a black hole of mass $\ga 10^8\msun$. It is reasonable to
suppose that the growth phases of these SMBH are observable as AGN.
But since the incidence of AGN among all galaxies is relatively low,
this must mean that the holes spend much of their time growing at the
maximum possible rate, i.e. that specified by the Eddington limit. In
this case the $M-\sigma$ relation constrains their growth, and we
should expect the black hole masses in AGN to lie {\it below} the
relation, with a galaxy nucleus ceasing to be active once its SMBH
reaches this mass. We are thus led to the conclusions that

\noindent(a) {\it AGN must radiate at close to the Eddington luminosity for
  much of their lives,} and

\noindent(b) {\it their black holes must be underweight, i.e below the $M -
  \sigma$ relation}.

Although these conclusions follow from accepting that observed AGN do
represent the growth phases of SMBH, neither of them is widely adopted
in practice.  This amounts either to simply ignoring the argument, or
to tacitly assuming that SMBH growth only happens in AGN which are
somehow unobservable.  I shall argue below that this tacit assumption
cannot be correct.

\section{$M - \sigma$ as an upper mass limit}

A major reason for the reluctance in accepting conclusion (b) above
has probably been the tightness of the observed $M - \sigma$
relation. However Batcheldor (2010) has recently pointed out that the
SMBH masses populating the observed $M - \sigma$ relation are found by
using the velocities of stars within the SMBH sphere of influence
\begin{equation}
R_{\rm inf} = {2GM\over \sigma^2}.
\label{inf}
\end{equation}
Since one must resolve $R_{\rm inf}$ in order to get an unambiguous
dynamical mass estimate, measured values of $M$ correlate with
$\sigma$. Batcheldor (2010) shows that it is unlikely that one can
measure masses $M$ significantly below the value given by the observed
$M - \sigma$ relation. Accordingly we are at liberty to regard the
observed relation as an upper limit on the SMBH mass for a given
$\sigma$.

\section{SMBH Feedback and the $M - \sigma$ relation}

We have inferred that SMBH grow at the Eddington rate during active
phases.  The resulting massive outflows are observed (see Section
5). They offer a natural way of communicating some of the hole's
binding energy to the host galaxy, and thus establishing the $M -
\sigma$ relation as they shock against the host interstellar
medium. Whatever the Eddington ratio $\dot m = \dot M/\me$, the
accretion luminosity is always $\sim \le$, because radiation pressure
expels excess accretion at each radius to ensure that the local
Eddington limit is not exceeded. Shakura \& Sunyaev (1973) show for
example that in a disc geometry the luminosity limit is of order $\le
(1 + \ln \dot m)$, where $\le$ is the standard expression for the
spherical Eddington limit. As the outflow results from radiation
driving, there are two extreme possibilities (cf King, 2003; 2005;
2010).

For $\dot m \sim 1$ most photons in the AGN radiation field scatter
about once and transmit their total momentum $\le/c$ to the outflow.
Further, the outflow shocks against the host ISM are effectively
cooled by Compton cooling from the AGN radiation field. Thus only the
outflow ram pressure $\le/c$ is communicated to the host: this is
a momentum--driven outflow.  The outflow sweeps up the host ISM in a
shell, and solving the shell's equation of motion shows that this
recollapses unless the SMBH mass has reached the critical value
\begin{equation}
M_{\sigma}({\rm mom}) = {f_g\kappa\over \pi G^2}\sigma^4
\label{msigmom}
\end{equation}
(where $f_g \simeq 0.16$ is the cosmic baryon fraction wrt dark
matter).  If the hole mass is $\geq M_{\sigma}$ the shell can reach
significant radii within the host and presumably suppress central
accretion and SMBH growth.  A simple way of deducing this result is to
equate the thrust $\le/c$ to the weight of the swept--up gas shell.
For an isothermal density distribution the weight of a swept--up gas
shell turns out to be $W = 2f_g\sigma^4/G$, independent of radius $R$
(the swept--up mass goes as $R$, while its gravity $\sim GM(R)/R^2$
goes as $1/R$ -- see King, 2010). Hence setting $\le \simeq W$ gives
an estimate of the mass. (The solution of the shell's equation of
motion shows that the radiation thrust accelerates the outflow to a
fixed terminal speed, justifying the approximation; cf King, 2005.)

At the opposite extreme, for $\dot m >> 1$, multiple scattering within
the outflow means that almost all the total photon energy is given to
the outflow.  The outflow shock against the ISM is now not effectively
cooled, and the total energy rate $\le$ now does $P{\rm d}V$ work
against the weight of the swept--up interstellar gas. This is an
energy--driven outflow, as originally considered by Silk \& Rees
(1998). Equating the energy injection rate $\le$ to the rate of
working $W\sigma$ as the shell begins to move to large radii we find
\begin{equation}
M_{\sigma}({\rm en}) \simeq {f_g\kappa\over \pi G^2 c}\sigma^5.
\label{msigen}
\end{equation}

The relations (\ref{msigmom}, \ref{msigen}) have no free parameter.
Each is potentially self--consistent: the accretion rate is likely to
be close to Eddington for the high--mass momentum--driven sequence
(i.e.  $\dot m \simeq 1$), and significantly super--Eddington for the
low--mass energy--driven sequence (i.e. $\dot m >>1$) (cf King, 2010).
The momentum--driven value $M_{\sigma}({\rm mom})$ is very close to
the observed relation
\begin{equation}
M \simeq 2\times 10^8\msun\sigma_{200}^4.
\label{msigobs}
\end{equation}
(Ferrarese \& Merritt, 2000; Gebhardt et al., 2000).

In contrast, the relation (\ref{msigen}) is a factor $\sim \sigma/c
\sim 10^{-3}$ below the observed one, and would imply unobservably
small hole masses.  It is important to understand why this low--mass
sequence is apparently disfavoured, as if not, it would constitute a
bottleneck to SMBH growth, and make it hard to understand why SMBH
apparently do reach the higher--mass sequence (\ref{msigmom}).

The reason why the energy--driven sequence does not seem to occur in
nature appears to be that the very strong density contrast between the
shocked wind and the host interstellar medium in the energy--driven
case causes the bubble blown by the outflow to break up through the
Rayleigh--Taylor instability.  By contrast, in a momentum--driven
outflow the postshock density is far higher, making the corresponding
bubble Rayleigh--Taylor stable for masses near $M_{\sigma}$, and
potentially able to cut off accretion.  In this latter case, the speed
$v$ of the preshock outflow is given by the condition
\begin{equation}
\dot M v = {\le\over c} = \eta\me c
\end{equation}
as
\begin{equation}
v  = {\eta \over \dot m}c
\label{vmom}
\end{equation}
The continuity equation gives the preshock density as
\begin{equation}
\rho = {\dot M\over 4\pi b R^2v} = {\dot m^2\me\over 4\pi R^2b\eta c}
\label{cont}
\end{equation}
at radius $R$, where $b$ is the fractional solid angle of the outflow
(we shall see later that $b$ is of order unity).  As the outflow shock
is efficiently cooled, the density is strongly increased (by a factor
$\sim (v/\sigma)^2$) there. Thus the outflow density in contact with
the host ISM is
\begin{equation}
\rho_m \simeq {\eta \dot m^4\me c\over 4\pi R^2b\sigma^2}
\end{equation}
The ISM density is roughly the isothermal value
\begin{equation}
\rho_{\rm ISM} = {f_g\sigma^2\over 2\pi R^2G}
\end{equation}
so combining we find
\begin{equation}
{\rho_{\rm ISM}\over \rho_m} \simeq {b\over 2\dot m^4}{M_{\sigma}({\rm
    mom})\over M}.
\end{equation}
Thus the bubble is Rayleigh--Taylor stable, and can propagate outwards
and suppress accretion provided that
\begin{equation}
M \ga {b\over 2\dot m^4}M_{\sigma}({\rm mom})
\label{m}
\end{equation}
Since $b \sim \dot m \sim 1$, we see that central accretion and SMBH
growth is likely to be suppressed for $M \sim M_{\sigma}({\rm
  mom})$. (Note that accretion of gas within the black hole sphere of
influence can continue for a time after the outflow bubble begins to
propagate to significant radii.) This suggests that SMBH masses in AGN
are likely to be only a factor of a few below the $M - \sigma$ limit
(\ref{msigmom}), as they can grow only when the externally imposed
accretion rate is not far above the Eddington value and so potentially
spend longer at such masses.

For an energy--driven outflow, the preshock outflow velocity follows from
\begin{equation}
{1\over 2}\dot Mv^2 = \le = \eta \me c^2
\end{equation}
as 
\begin{equation}
v = \left({2\eta\over \dot m}\right)^{1/2}.
\end{equation}
With this change, and compression only by a factor 4 in the outflow shock 
(since cooling is negligible), the SMBH growth suppression 
condition corresponding to (\ref{m}) becomes
\begin{equation}
M > {b\over 2\dot m^{3/2}}\left({c\over
  \sigma}\right)^3M_{\sigma}({\rm en}) \sim 10^9{b\over\dot
  m^{3/2}}M_{\sigma}({\rm en})
\end{equation}
which is clearly impossible to satisfy for realistic parameters. Thus
an energy--driven bubble is never Rayleigh--Taylor stable and cannot
halt SMBH growth. We conclude that SMBH growth is not halted at the
energy--driven sequence (\ref{msigen}), and given an adequate mass
supply can continue all the way to the momentum--driven limit
(\ref{msigmom}).

\section{Observed Outflows}

We have seen that SMBH feedback occurs through massive outflows
carrying the Eddington momentum. These outflows must have speeds $v
\simeq 0.1c$ (cf eqn \ref{vmom}). Moreover since the momentum outflow
rate is specified by the quantity $\rho R^2v$ (cf eqn \ref{cont}), the
ionization parameter
\begin{equation}
\xi = {L_i\over NR^2} \simeq {l_i\le m_p\over \rho R^2}
\end{equation}
is also specified (here $L_i = l_i\le$ is the ionizing luminosity of
the AGN, with $l_i$ given by the spectral shape and ionization
threshold). King (2010) shows that this condition requires that the
outflows should have ionization parameters
\begin{equation}
\xi = 3\times 10^4\eta_{0.1}^2l_2\dot m^{-2},
\label{ion2}
\end{equation}
where $l_2 = l_i/10^{-2}$ and $\eta_{0.1} = \eta/0.1$, and thus show
lines in the X--ray region of the spectrum. This is indeed what is
observed (Pounds et al., 2003a, b; O'Brien et al, 2005).  This
reasoning shows that
{\it any outflow with velocities $\sim 0.1c$ seen in X--ray
  lines carries mass and momentum rates of order $\me, \le/c$}.

The papers by Tombesi et al. (2010a,b) show that outflows with these
properties are extremely common.  They are detected in a significant
fraction ($>35\%$) of a sample of more than 50 local AGN. This must
mean first that the solid angle factor $b$ cannot be small (Tombesi et
al. deduce $b\ga 0.6$), and second, that a large number of local AGN
have undergone or continue to undergo episodes of Eddington accretion.

We get more information from the hydrogen column
densities, which are observed to lie in the range 
$10^{22}~{\rm cm^{-2}} < N_H < 10^{24}~{\rm cm^{-2}}$. Using eqn. (\ref{cont})
we find 
\begin{equation}
N_H = \int_{R_{\rm in}}^{\infty}{\rho\over m_p}{\rm d}R = {\dot
  m^2\me\over 4\pi b\eta c m_p R_{\rm in}} \simeq {3\times 10^{37}\dot
  m^3M_8\over b\eta_{0.1}^2 R_{\rm in}}~{\rm cm}^{-2}
\label{NH}
\end{equation}
where $R_{\rm in}$ is the inner radius of the flow. For a continuing
outflow this would be of order 100 Schwarzschild radii (corresponding
to the escape speed $\sim 0.1c$).  It is difficult to detect the
corresponding very high column densities $\sim 10^{24}~{\rm cm^{-2}}$
as the gas is likely to be fully ionized.  The highest observed
columns presumably correspond to continuing steady outflows which have
recombined at some distance from the black hole. Lower columns may
reveal cases where the outflow is episodic, and last stopped at a time
$t_{\rm off}$ before the observation.  In this case $R_{\rm in}$ takes
a larger value $\simeq vt_{\rm off}$.  Using (\ref{NH}) we get
\begin{equation}
t_{\rm off} = 0.2{\dot m^3M_8\over b\eta_{0.1}^2N_{23}}~{\rm yr},
\label{toff}
\end{equation}
where $M_8 = M/10^8\msun, N_{23} = N_H/(10^{23}~{\rm cm}^{-2})$.  Thus
even the lowest observed columns ($\sim 10^{22}~{\rm cm^{-2}}$)
correspond to outflows which switched off only 2 years ago. This
suggests that outflows are even more common than one might expect from
the simplest interpretation of the results of Tombesi et al (2010a,
b).

\subsection{Eddington or not?}

Tombesi et al.  (2010b) consider three sources in detail (3C 111, 3C
120, 3C 390.3) and suggest that their luminosities are below
$\le$. However this procedure uses black hole masses based on
assumptions that either the $M - \sigma$ relation, or the SMBH--bulge
mass relation, hold for {\it all} SMBH, at least in a statistical
sense. (The $M-\sigma$ relation is used to calibrate the reverberation
masses for 3C 120, 3C 390.3; Peterson et al., 2004, and the SMBH mass
for 3C 111 uses the stellar bulge luminosity; Marchesini et al.,
2004.)  We have seen that both relations are likely to give
overestimates of the SMBH mass, and hence overestimates of
$\le$. Figure 16 of Peterson et al., (2004) shows that a reduction by
a factor $\sim 3$ of SMBH masses estimated in this way would put most
of the sample of 35 reverberation--mapped AGN close to their Eddington
luminosities. In addition to this, estimating whether an AGN is near
$\le$ requires us to know not only its SMBH mass, but also its true
bolometric luminosity $L_{\rm bol}$, both to high accuracy. The latter
problem is unlikely to be easier than the former.

\section{Conclusions}

This paper has argued that the black hole mass is a factor of a few
below the $M - \sigma$ mass in active galaxies, and that a large
fraction of AGN are fed mass at a super--Eddington rate, accreting
just the Eddington value and expelling the excess. 

The first point follows from noting that SMBH growth towards the
momentum--driven limit (\ref{msigmom}) is inevitable given a
sufficient mass supply. In particular, energy--driven outflows are
Rayleigh--Taylor unstable, so the mass is not constrained by the
energy--driven limit (\ref{msigen}). Growth only slows when
momentum--driven outflows become Rayleigh--Taylor stable, i.e. when the
black hole is a factor of a few below the $M - \sigma$ value. So SMBH
masses in AGN are likely to be below, but fairly close to, this
critical value. This agrees with the suggestion by Batcheldor (2010)
that the $M - \sigma$ relation is an upper limit to SMBH masses.

The idea that AGN regularly reach $\le$ follows naturally from noting
that SMBH have to grow rapidly to reach the masses specified by the
Soltan relation. It is consistent with the first proposition above
(low SMBH masses): Eddington ratios for observed AGN must be higher
than previously estimated if their black holes lie below the $M -
\sigma$ relation rather than on it, as is sometimes assumed.

The strongest evidence for Eddington accretion comes from the papers
by Tombesi et al. (2010a, b), which show that a large fraction of
nearby AGN have outflow with velocities $\sim 0.1c$ and ionization
parameters $\xi \sim 10^4 - 10^5$, as expected. At face value this
suggests that a large fraction of local AGN are fed at
super--Eddington rates, and I have argued that it is difficult to
avoid this conclusion.

\section{Acknowledgments}

I thank Brad Peterson, Ken Pounds, James Reeves, and Marianne
Vestergaard for illuminating discussions, and the referee for a very
thorough reading of the paper. Theoretical astrophysics
research at Leicester is supported by an STFC rolling grant.


\begin{thebibliography}{}

\bibitem[\protect\citeauthoryear{Batcheldor}{2010}]{2010ApJ...711L.108B} 
Batcheldor D., 2010, ApJ, 711, L108 

\bibitem[\protect\citeauthoryear{Ferrarese \&
    Merritt}{2000}]{2000ApJ...539L...9F} Ferrarese L., Merritt D.,
  2000, ApJ, 539, L9

\bibitem[\protect\citeauthoryear{Gebhardt et 
al.}{2000}]{2000ApJ...539L..13G} Gebhardt K., et al., 2000, ApJ, 539, L13 

\bibitem[\protect\citeauthoryear{H{\"a}ring \&
    Rix}{2004}]{2004ApJ...604L..89H} H{\"a}ring N., Rix H.-W., 2004,
  ApJ, 604, L89

\bibitem[\protect\citeauthoryear{King}{2003}]{2003ApJ...596L..27K} King A., 
2003, ApJ, 596, L27 

\bibitem[\protect\citeauthoryear{King}{2005}]{2005ApJ...635L.121K} King A., 
2005, ApJ, 635, L121 

\bibitem[\protect\citeauthoryear{King}{2010}]{2010MNRAS.402.1516K} King 
A.~R., 2010, MNRAS, 402, 1516 

\bibitem[\protect\citeauthoryear{Marchesini, Celotti, \&
    Ferrarese}{2004}]{2004MNRAS.351..733M} Marchesini D., Celotti A.,
  Ferrarese L., 2004, MNRAS, 351, 733


\bibitem[\protect\citeauthoryear{O'Brien et 
al.}{2005}]{2005MNRAS.360L..25O} O'Brien P.~T., Reeves J.~N., Simpson C., 
Ward M.~J., 2005, MNRAS, 360, L25 

\bibitem[\protect\citeauthoryear{Peterson et 
al.}{2004}]{2004ApJ...613..682P} Peterson B.~M., et al., 2004, ApJ, 613, 
682 

\bibitem[\protect\citeauthoryear{Pounds et al.}{2003a}]{2003MNRAS.345..705P} 
Pounds K.~A., Reeves J.~N., King A.~R., Page K.~L., O'Brien P.~T., Turner 
M.~J.~L., 2003, MNRAS, 345, 705 

\bibitem[\protect\citeauthoryear{Pounds et al.}{2003b}]{2003MNRAS.346.1025P} 
Pounds K.~A., King A.~R., Page K.~L., O'Brien P.~T., 2003, MNRAS, 346, 1025 

\bibitem[\protect\citeauthoryear{Shakura \&
    Sunyaev}{1973}]{1973A&A....24..337S} Shakura N.~I., Sunyaev R.~A.,
  1973, A\&A, 24, 337



\bibitem[\protect\citeauthoryear{Soltan}{1982}]{1982MNRAS.200..115S} Soltan 
A., 1982, MNRAS, 200, 115 

\bibitem[\protect\citeauthoryear{Tombesi et 
al.}{2010a}]{2010arXiv1006.2858T} Tombesi F., Cappi M., Reeves J.~N., 
Palumbo G.~G.~C., Yaqoob T., Braito V., Dadina M., 2010, arXiv, 
arXiv:1006.2858, A\&A, in press

\bibitem[\protect\citeauthoryear{Tombesi et 
al.}{2010}]{2010ApJ...719..700T} Tombesi F., Sambruna R.~M., Reeves J.~N., 
Braito V., Ballo L., Gofford J., Cappi M., Mushotzky R.~F., 2010, ApJ, 719, 
700 



\end{thebibliography}
\end{document}